\documentclass[12pt]{JHEP3}
\usepackage[english]{babel}
\usepackage{amsmath,amssymb}
\usepackage[dvips]{graphicx}
\usepackage{rotating}
\usepackage{cite}
\usepackage{wrapfig}

\preprint{}
\title{Soliton pair creation in classical wave scattering}
\author{S.V. Demidov, D.G. Levkov\\
  Institute for Nuclear Research of the Russian Academy of
  Sciences,\\ 60th October Anniversary Prospect 7a, 117312, Moscow,
  Russia\\ E-mail: \email{demidov@ms2.inr.ac.ru},
  \email{levkov@ms2.inr.ac.ru}}
\abstract{We study classical production of soliton--antisoliton pairs
  from colliding wave packets in $(1+1)$--dimensional scalar field model.
  Wave packets represent multiparticle states in 
  quantum theory; we characterize them by energy $E$ and particle
  number $N$. Sampling stochastically over the forms of wave packets,
  we find the entire region in $(E,N)$ plane which corresponds to
  classical creation of soliton pairs. Particle number is
  parametrically large within this region meaning that the probability
  of soliton--antisoliton pair production in few--particle collisions is 
  exponentially   suppressed.}
 
\keywords{Solitons Monopoles and Instantons, Nonperturbative Effects}

\begin{document}
\section{Introduction}
\label{sec:intro}
Wonders related to classical dynamics of solitons in non--integrable
models surprised theorists for
decades~\cite{Makhankov,soliton_chaos,Belova,Weinberg}. 
Intriguing long--living bound states of solitons and 
antisolitons --- oscillons --- are found in a variety of
models~\cite{oscillon0,oscillon_SU2,oscillons_explanation,%
oscillon_vortex,electroweak_oscillon}.
Another interesting example is kink--antikink annihilation
in $(1+1)$--dimensional $\phi^4$ theory which displays chaotic
behavior~\cite{resonance_JETP,resonance0,fractal}. Most of these
phenomena are explained qualitatively by reducing the infinite
number of degrees of freedom in field theory to a few collective
coordinates~\cite{Manton1,Manton2,Manton_kinks}. Then, mechanical
motion along the collective coordinates shows whether soliton
evolution is regular or chaotic.

\begin{sloppy}
Recently~\cite{kinks_particles,Shnir} a question of kink--antikink
pair production in classical wavepacket scattering was
addressed\footnote{Related yet different problem is creation of
  kink--antikink pairs from wave packets in the background of
  preexisting kink~\cite{kink_to_3kinks}.}, cf.
Refs.~\cite{abelian_collisions,phi4_collisions,SU2_collisions0,%
SU2_collisions}. The 
interest to this question stems from the fact that, within the
semiclassical approach, wave packets describe multiparticle
states in quantum theory. Studying kink--antikink creation from
wave packets one learns a lot about the quantum counterpart process:
production of nonperturbative kink states in multiparticle
collisions. The prospect of Refs.~\cite{kinks_particles,Shnir} was to 
describe a class of multiparticle states leading to classical
formation of kinks.

\end{sloppy}
Due to essential nonlinearity of classical field equations the process
of kink--antikink creation cannot be described analytically\footnote{In
  particular, collective coordinates cannot be introduced since
  solitons are absent in the beginning of the process.} and
one has to rely on numerical methods. A difficulty, however, is
related to the space of initial Cauchy data which is
infinite--dimensional  in field theory. Because of this difficulty the
analysis of Refs.~\cite{kinks_particles,Shnir} was limited to
a few--parametric families of initial data.

In this paper we explore the entire space of classical solutions
describing soliton--antisoliton pair creation from wave packets. To
this end we sample stochastically over the sets of Cauchy data and
obtain large ensemble of solutions, cf. Refs.~\cite{Rebbi1,Rebbi2}. We
select solutions evolving between free wave packets and
soliton--antisoliton pair and compute the energies $E$ and particle
numbers $N$ of the respective initial states. In this way we obtain
the region in $(E,N)$ plane corresponding to classical creation of
solitons. We are particularly interested in solutions from this region
with the smallest $N$.

The model we consider is somewhat different from the standard $\phi^4$
theory used in Refs.~\cite{kinks_particles,Shnir}. We do
study evolution of a scalar field in $(1+1)$ dimensions but choose
nonstandard potential $V(\phi)$ shown in Fig.~\ref{fig:V}a, solid
line. 
\FIGURE{
\centerline{\includegraphics[width=0.5\textwidth]{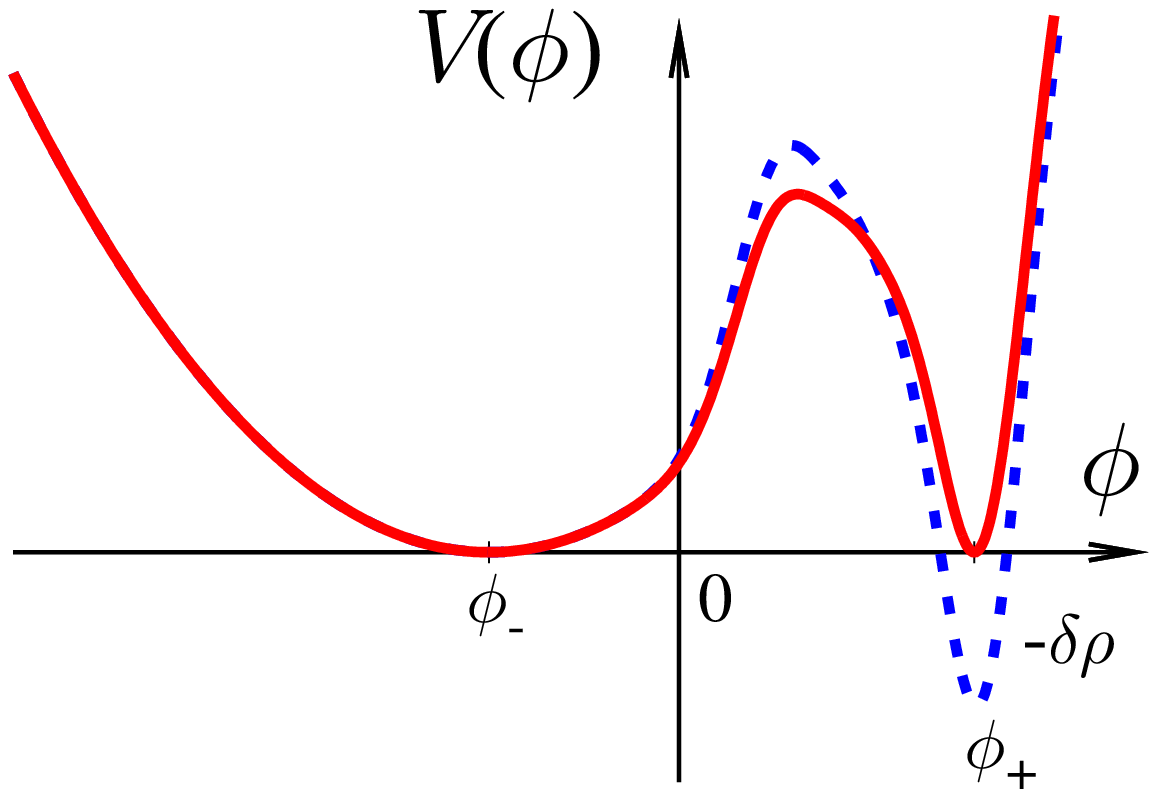}\hspace{5mm}
\includegraphics[width=0.42\textwidth]{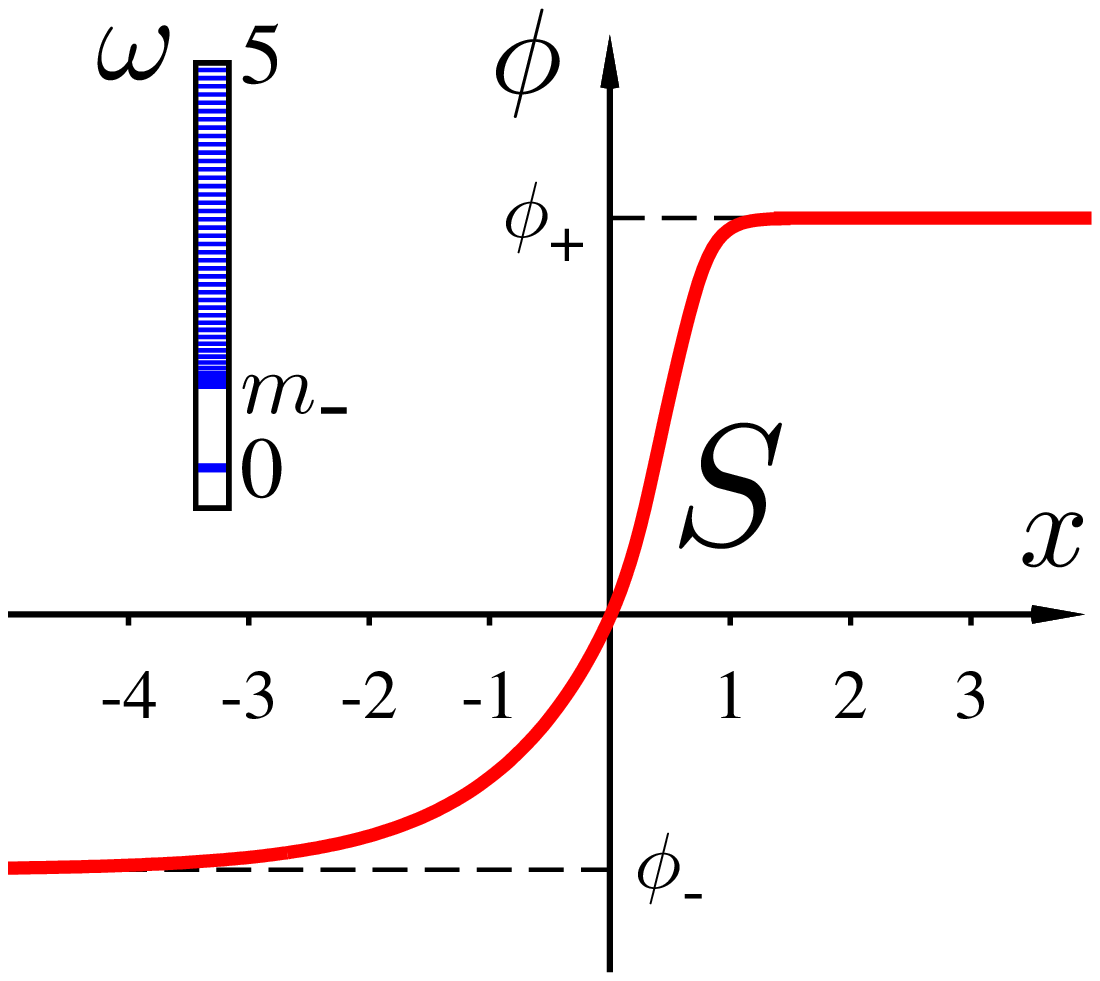}}
\centerline{\hspace{1.2cm}(a) \hspace{6.7cm}(b)}
\caption{\label{fig:V}(a) Potential $V(\phi)$. (b) Soliton
  and its spectrum.}
}
The reason for the unusual choice is chaos in kink--antikink
scattering in $\phi^4$ theory: it would be a venture to 
try applying new method in a potentially nontrivial chaotic
model. We will comment on generalizations of our technique in the
Discussion section.

With the above set of classical solutions we test the method
of Ref.~\cite{DL} where classically forbidden production of kinklike
solitons in the same model was studied. Namely, we compare the
boundary of the ``classically allowed'' region in $(E,N)$ plane with
the same boundary obtained in Ref.~\cite{DL} from the classically
forbidden side. Coincidence of the two results justifies both
calculations.

The paper is organized as follows. We introduce the model in
Sec.~\ref{sec:model} and explain the stochastic sampling technique in
Sec.~\ref{sec:method}. In Sec.~\ref{sec:results} we present numerical
results which confirm, in particular, results of Ref.~\cite{DL}. We
conclude and generalize in Sec.~\ref{sec:discussion}. 

\section{The model}
\label{sec:model}
The action of the model is
\begin{equation}
\label{eq:11}
S = \frac{1}{g^2} \int dt \, dx\, \left[(\partial_\mu \phi)^2/2  -
  V(\phi)\right]\;, 
\end{equation}
where $\phi(t,x)$ is the scalar field; 
semiclassical parameter $g$ does not enter the classical field
equation
\begin{equation}
\label{eq:1}
\left(\partial_t^2  - \partial_x^2\right) \phi = - \partial
V(\phi)/\partial \phi\;.
\end{equation}
We assume that the potential $V(\phi)$ has a pair of
degenerate minima $\phi_-$ and $\phi_+$. Then there exists a
static solution of Eq.~(\ref{eq:1}) --- topological kinklike soliton
$\phi_S(x)$ shown in Fig.~\ref{fig:V}b. Antisoliton solution
$\phi_A(x)$ is obtained from $\phi_S(x)$ by spatial reflection,
$\phi_A(x) = \phi_S(-x)$. 

We consider classical evolutions of $\phi(t,x)$ between free
wave packets in the vacuum $\phi_-$ and
configurations containing soliton--antisoliton pair. Initial and final 
states of the process are shown schematically in
Fig.~\ref{fig:process}.
\FIGURE{
\centerline{\includegraphics[width=0.5\textwidth]{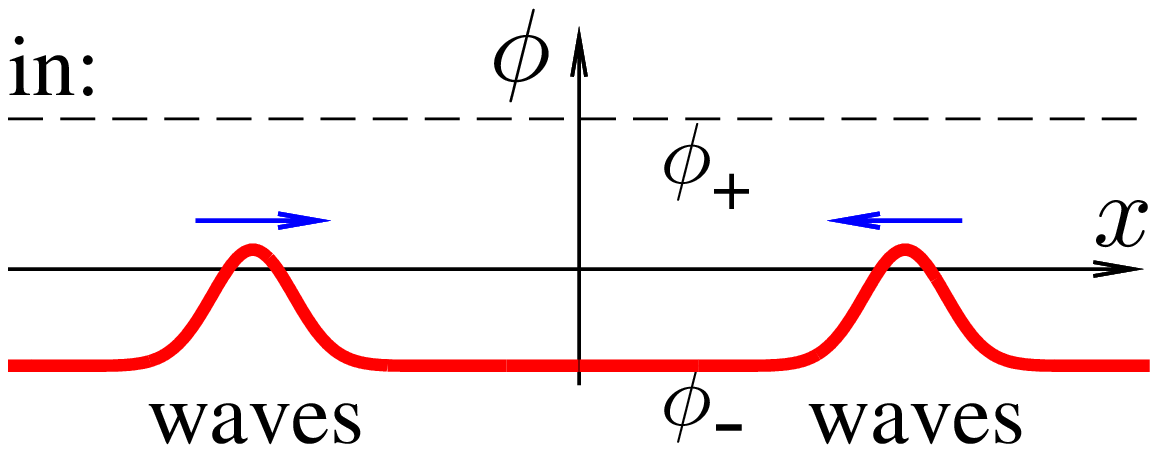}
\includegraphics[width=0.5\textwidth]{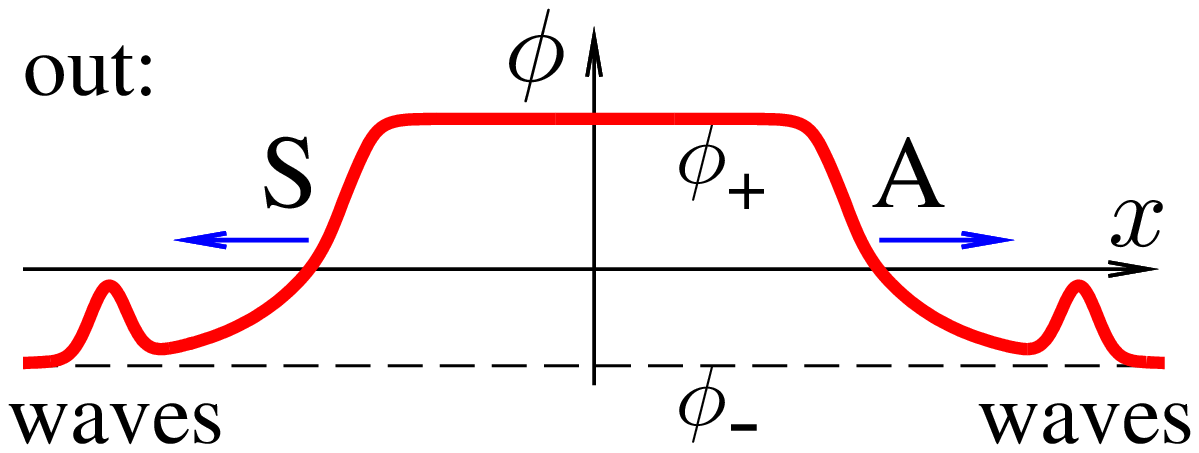}}
\caption{\label{fig:process}Classical formation of
  soliton--antisoliton pair.}
}
We restrict attention to $P$--symmetric solutions,
$\phi(t,x)=\phi(t,-x)$. This is natural since soliton and antisoliton
are symmetric with respect to each other.

In what follows we solve Eq.~(\ref{eq:1}) numerically. To this end we
introduce a uniform spatial lattice $\{x_{i}\}$, $i=- N_x,\dots
N_x$ of extent $-L_x \le x_{i} \le L_x$. At lattice edges $x=\pm
L_x$ we impose energy--conserving Neumann boundary conditions
$\partial_x \phi =0$. We also introduce a uniform time step $\Delta
t$. Typically, $L_x =  15$, $N_x = 400$, $\Delta t =
0.03$. Discretization of Eq.~(\ref{eq:1}) is standard
second--order\footnote{One 
  changes $\partial_x^2\phi(x_i)$ to $(\phi_{i+1} + 
  \phi_{i-1} - 2\phi_i)/\Delta x^2$, where $\phi_i = \phi(x_i)$. The
  time derivative $\partial_t^2 \phi$ is discretized in the same
  way.}.
We take advantage of the reflection symmetry $x\to -x$ and use only
one half of the spatial lattice.

We consider the potential
\begin{equation}
\label{eq:2}
V(\phi) = \frac{1}{2}(\phi+1)^2\left[1 - v\,
  W\left(\frac{\phi-1}{a}\right)\right]\;,
\end{equation}
where dimensionless units are introduced; $W(x) = \mathrm{e}^{-x^2}( x
+ x^3 + x^5)$, $a=0.4$. The value of $v$ is chosen to equate the
energy densities of the vacua, $v\approx 0.75$. The potential
(\ref{eq:2}) is depicted in Fig.~\ref{fig:V}a, solid line. 
We denote the masses of linear excitations in the vacua $\phi_-$ and $\phi_+$
by $m_-$ and $m_+$, respectively.

Our choice of the potential is motivated in two ways. First, we have
already mentioned that kink dynamics in the standard
$\phi^4$ theory is
chaotic~\cite{resonance_JETP,resonance0,fractal}. The source of 
chaos hides in the spectrum of linear perturbations around the
$\phi^4$ kink. The latter contains {\it two} localized modes:
zero mode due to spatial translations and first excited mode
representing kink periodic pulsations. Localized modes accumulate
energy during kink evolution which is thus described by two collective
coordinates. Mechanical model for these coordinates is
chaotic~\cite{fractal}, just like the majority of two--dimensional
mechanical models. 

We get rid of the chaos by choosing the potential (\ref{eq:2}) where the
spectrum of linear perturbations around the soliton contains only one
localized mode. Due to this property soliton motion is described
by one--dimensional mechanical system which cannot be chaotic. 

Let us compute the spectrum of the soliton in the model
(\ref{eq:2}). Consider small perturbations $\phi -
\phi_S(x) = \delta\phi (x)\cdot\mathrm{e}^{\pm i\omega t}$ in the
background of the soliton. Equation~(\ref{eq:1}) implies,
\begin{equation}
\label{eq:3}
\left[- \partial_x^2 + U(x)\right] \delta
\phi(x) = \omega^2 \delta \phi(x)\;,
\end{equation}
where $U(x) = V''(\phi_S(x))$ and nonlinear terms in $\delta\phi$
are neglected. Discretization turns the differential operator in 
Eq.~(\ref{eq:3}) into  a symmetric 
$(2N_x+1)\times (2N_x+1)$ matrix; we compute the eigenvalues
$\{\omega_k^{(S)}\}$ of this matrix by the standard method of
singular value decomposition. Several lower eigenvalues are shown in the
inset in Fig.~\ref{fig:V}b. One sees no localized modes between zero
mode and continuum $\omega^{(S)} > m_-$.

Another, unrelated to the soliton spectrum, mechanism of chaos was
proposed recently in Ref.~\cite{Dorey:2011yw}. This mechanism works
under condition $m_+ < m_-$ which is not met in our model.

The second reason for the choice~(\ref{eq:2}) is linearization of 
classical solutions at large negative times. Interaction terms
should be negligible in the initial part of the
classical evolution; otherwise initial wave packets cannot be
associated with the perturbative Fock states. However,
$(1+1)$--dimensional solutions linearize slowly due to wave 
dispersion. Brute force linearization would require large
lattice which is a challenge 
for the numerical method. Our model is  specifically designed to
overcome this difficulty. At $a\ll 1$ the potential~(\ref{eq:2}) is
quadratic everywhere except for the small region $\phi\approx 1$. Wave
packets move freely in this potential if their tops are away
from $\phi=1$, see Fig.~\ref{fig:process}. After collision the wave
packets add up coherently and hit the interaction region $\phi\approx
1$. Below we find that all classical solutions of interest behave in
the described way. We use $a=0.4$ which is small enough to provide,
for the chosen lattice size, linearization at the level of~1\%. 

It is worth noting that the problem with slow linearization is absent
in multidimensional theories because amplitudes of spherical waves
in $D>2$ decay at power laws with distance. In general 
$(1+1)$--dimensional model linearization can be achieved artificially
by switching off the interaction terms of the potential at
$|x| > L_{int}$. This corresponds to a physical setup where
interaction takes place in a sample of length $2 L_{int}$.

We compute the energy $E$ and particle number $N$ of the initial
wave packets in the following way. Since the wave packets move freely
in the vacuum $\phi_{-}$,
\begin{equation}
\label{eq:4}
\phi(t,x) \to \phi_- + \sqrt{\frac{2}{\pi}}\int_0^{\infty}
\frac{dk}{\sqrt{2\omega_k}} \,\cos(kx)\left[a_k
  \mathrm{e}^{-i\omega_k t} + 
  a_k^* \mathrm{e}^{i\omega_k t}\right]
\qquad{as} \;\; t\to -\infty\;,
\end{equation}
where we took into account the reflection symmetry and introduced
the amplitudes $a_k$; ${\omega_k^2 = k^2 + m^2_-}$. Given the
representation (\ref{eq:4}), one calculates $E$ and $N$ by the
standard formulas, 
\begin{equation}
\label{eq:5}
E = \frac2{g^2}\int_0^\infty dk \,\omega_k  |a_k|^2\;,\qquad \qquad
N = \frac2{g^2}\int_0^\infty dk \,|a_k|^2\;.
\end{equation}
Expression for $N$ can be thought of as a sum of mode occupation
numbers $n_k=|a_k|^2$, where the latter are defined as ratios of 
mode energies $\omega_k |a_k|^2$ and energy quanta\footnote{In our
  units $\hbar=1$.} $\omega_k$. Note that the energy $E$ is conserved;
it can be calculated at arbitrary moment of classical evolution as
\begin{equation}
\label{eq:8}
E=\frac1{2g^2}\int dx \left[(\partial_t \phi)^2 + (\partial_x\phi)^2 +
  2V(\phi)\right]\;.
\end{equation}
In the case of free evolution this expression coincides with the first of
Eqs.~(\ref{eq:5}). Needless to say that Eqs.~(\ref{eq:5}) can be used
only in the linear regime; this is the practical reason for continuing
solutions back in time until Eq.~(\ref{eq:4}) holds. Below we check
the linearity of classical solutions by comparing their exact and linear
energies, Eqs.~(\ref{eq:8}) and (\ref{eq:5}). We
characterize classical solutions by points in  $(E,N)$ plane.

Expressions (\ref{eq:4}), (\ref{eq:5}) are naturally generalized to
the lattice system.\footnote{Discretization of Eq.~(\ref{eq:8}) is
  standard second--order.} One solves numerically the eigenvalue problem 
(\ref{eq:3}), where $U(x) = m_-^2$, and finds the spectrum
$\{\delta\phi_k(x),\,\omega_k\}$ of linear excitations above the vacuum
$\phi_-$. In this way one obtains lattice analogs of the standing waves
$\cos(kx)$ and frequencies $\omega_k=\sqrt{k^2+m_-^2}$. Arbitrary
linear evolution in the vacuum $\phi_-$ has the form
\begin{equation}
\label{eq:6}
\phi(t,x) = \phi_- + \sum_k \delta\phi_k(x)
\left[a_k  \mathrm{e}^{-i\omega_k t} +
  a_k^* \mathrm{e}^{i\omega_k t }\right]\;,
\end{equation}
cf. Eq.~(\ref{eq:4}), where we used the eigenmode basis with 
normalization
\begin{equation}
\label{eq:6_5}
\sum_i \Delta x \,\delta\phi_k(x_i) \delta\phi_{k'}(x_i)  = \delta_{k,k'}
/\omega_k\;,\;\;
\Delta x = x_{i+1}-x_{i}. 
\end{equation}
One extracts the amplitudes $a_k$ from the classical solution
$\phi(t,x)$ at large negative $t$ by decomposing $\phi(t,x)$,
$\partial_t\phi(t,x)$ in the basis of $\delta\phi_k(x)$ and comparing the
coefficients of decomposition with Eq.~(\ref{eq:6}). Summing up the
energies and occupation numbers of different modes, one obtains
\begin{equation}
\label{eq:7}
E = \frac{2}{g^2}\sum_k \omega_k |a_k|^2 \;, \qquad \qquad 
N = \frac{2}{g^2}\sum_k |a_k|^2\;,
\end{equation}
where Eq.~(\ref{eq:6_5}) is taken into account.

\section{The method}
\label{sec:method}

\subsection{Modification of the potential}
\label{sec:modification}
It is difficult to select solutions containing 
soliton--antisoliton pairs in the infinite
future. On the one hand, numerical methods do not allow us to extend
$\phi(t,x)$ all the way to $t\to +\infty$. On the other hand,
soliton and antisoliton attract; taken at rest, they accelerate towards
each  other and annihilate classically  into a collection of
waves. Thus, we never can be sure that $\phi(t,x)$ contains solitons
at $t\to +\infty$, even if lumps similar to soliton--antisoliton pairs
are present at finite times. We solve this difficulty by changing
the value of $v$ in Eq.~(\ref{eq:2}) and thus adding small negative
energy density $(-\delta \rho)$ to the vacuum $\phi_+$, see
Fig.~\ref{fig:V}a, dashed line. This turns soliton--antisoliton pair
into a bubble of true vacuum $\phi_+$ inside the false vacuum
$\phi_-$~\cite{false,false1,false2}. Large bubbles 
expand at $\delta\rho>0$ since attraction between the solitons in this
case is surmounted by the constant pressure 
$\delta\rho$ inside the bubble. Thus, at $\delta\rho > 0$ we
simply look whether solution $\phi(t,x)$ contains large bubbles at
finite $t$. In the end of calculation, however, we have to
consider the limit $\delta\rho \to 0$.

We remark that solutions containing soliton--antisoliton pairs at
$t\to +\infty$ can be identified by other methods. Our way,
besides being particularly simple, has the following advantage: at
$\delta\rho >0$ there exists a critical bubble~\cite{false} ---
unstable static solution $\phi_{cb}(x)$ lying on top of  the potential
barrier between the true and false vacua. Given the critical bubble,
one easily constructs classical evolutions between the 
vacua. Indeed, in the critical bubble attraction between the soliton
and antisoliton is equal to repulsion due to $\delta\rho$. Being
perturbed, it either starts expanding or collapses forming a
collection of waves in the vacuum $\phi_-$. Thus, adding small
perturbation to the critical bubble and solving classical equations of
motion forward and backward in time, one obtains the classical
solutions of interest. Critical bubble at $\delta \rho=0.4$ is depicted
in Fig.~\ref{fig:sphaleron}a.
\FIGURE{
\unitlength=0.01\textwidth
\begin{picture}(100,55)
\put(-4.8,4){\includegraphics[width=0.55\textwidth]{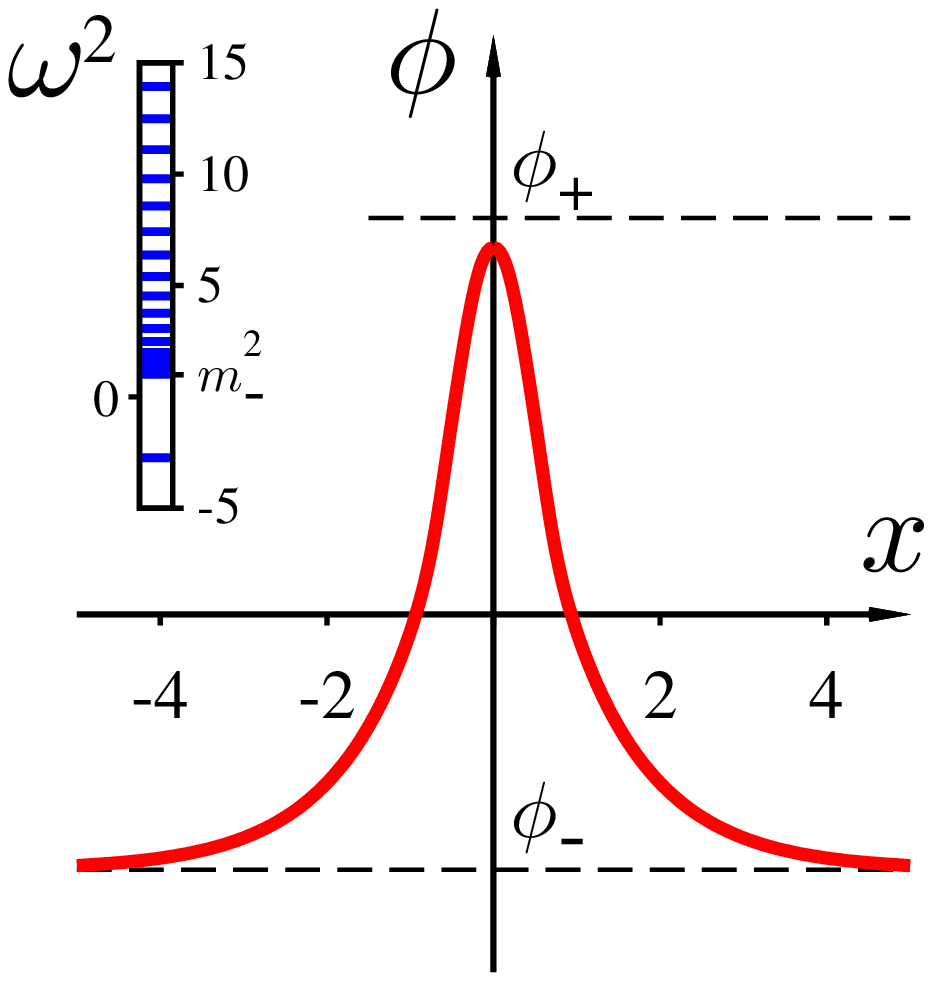}}
\put(52,20){\includegraphics[width=0.47\textwidth]{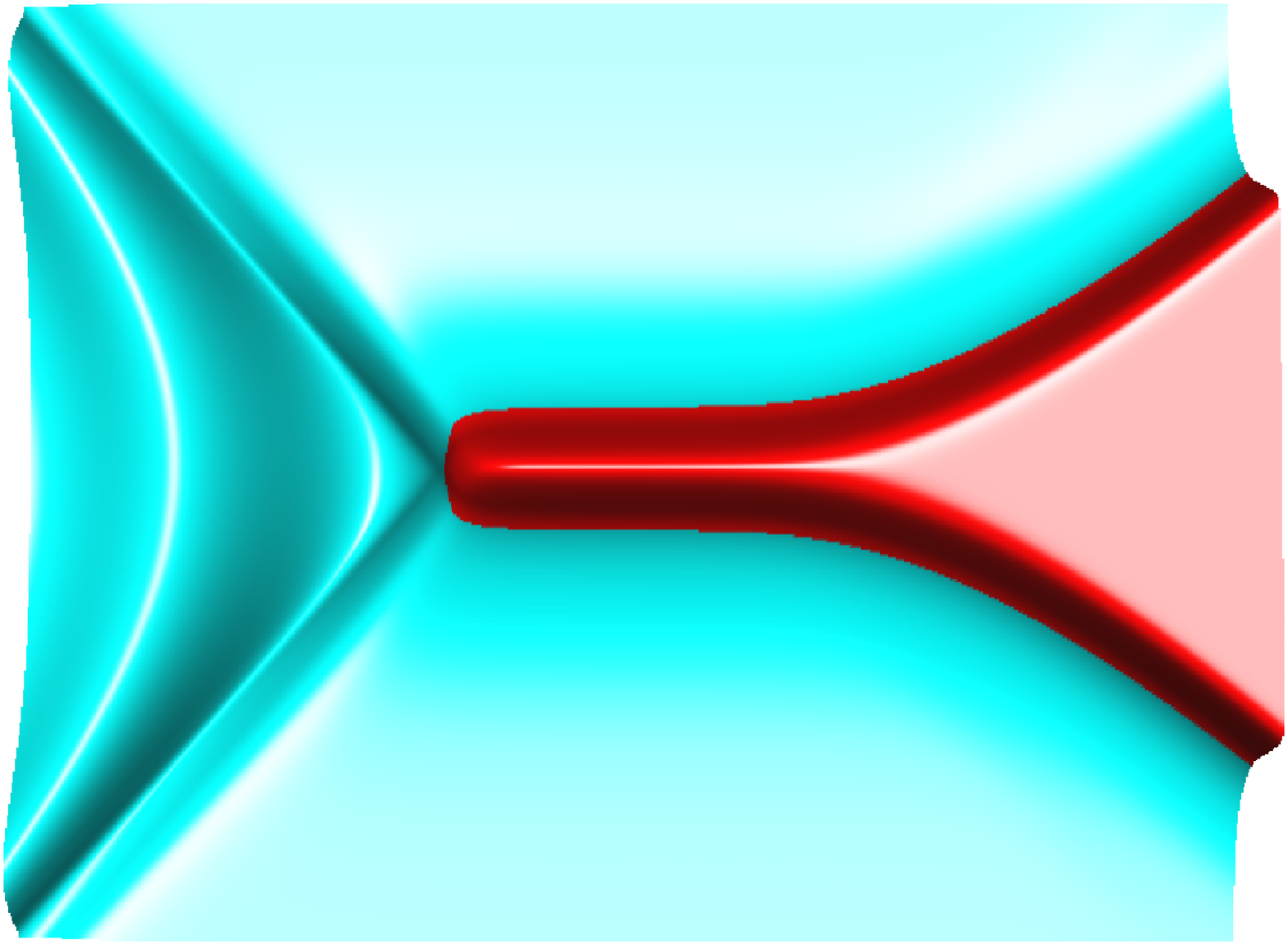}}
\put(46,-1){\includegraphics[width=0.537\textwidth]{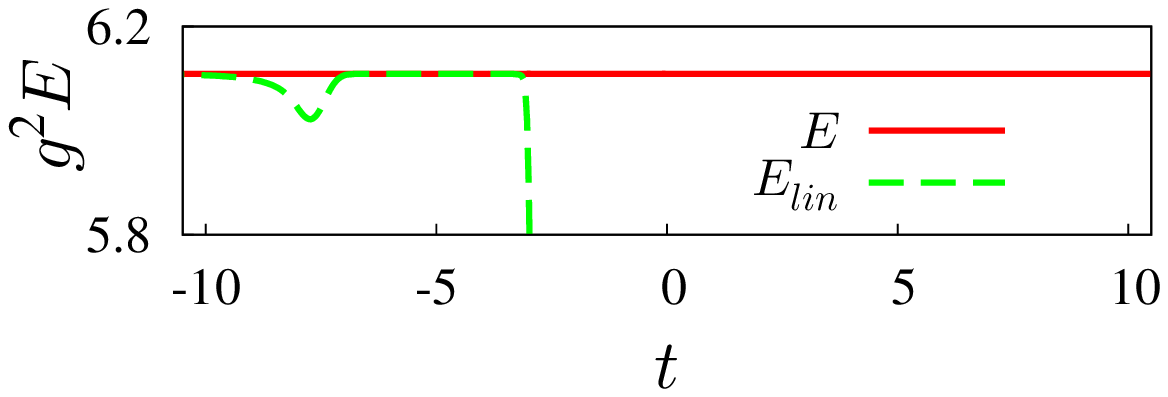}}
\put(76,16){\large $t$}
\put(74.9,19){\footnotesize $0$}
\put(62.7,19){\footnotesize $-5$}
\put(53,19){\footnotesize $-10$}
\put(84.7,19){\footnotesize $5$}
\put(94,19){\footnotesize $10$}
\put(50.2,24){\footnotesize $-6$}
\put(50.2,28.5){\footnotesize $-4$}
\put(50.2,33){\footnotesize $-2$}
\put(52,37.4){\footnotesize $0$}
\put(52,41.9){\footnotesize $2$}
\put(52,46.4){\footnotesize $4$}
\put(52,50.8){\footnotesize $6$}
\put(47,37){\large\begin{sideways}$x$\end{sideways}}
\end{picture}
\centerline{(a) \hspace{7.2cm} (b)\hspace{3mm}}
\caption{\label{fig:sphaleron}(a) Critical bubble and (b) solution
  $\phi(t,x)$ describing its classical decay; ${\delta\rho = 0.4}$. Red
  (dark gray) and blue 
  (light gray) colors in Fig.~\ref{fig:sphaleron}b mark regions with
  $\phi>0$ and $\phi<0$, respectively.}  
}

Let us obtain a particular solution describing creation of expanding
bubble from wave packets at $\delta\rho>0$. 
We solve numerically Eq.~(\ref{eq:3}) with $U(x) = V''(\phi_{cb}(x))$
and find the spectrum of linear perturbations $\{\delta
\phi_k^{(cb)}(x)\,, \omega_k^{(cb)}\}$ around the critical
bubble. This spectrum is shown in the inset 
in Fig.~\ref{fig:sphaleron}a; it contains precisely one negative mode
$\delta \phi_{neg}(x)$, $\omega_{neg}^2 < 0$ due to changes in the
bubble size. The latter mode describes decay of the critical bubble,
\begin{equation}
\label{eq:10}
\phi(t,x) \approx \phi_{cb}(x) + B_{neg}\,
\delta\phi_{neg}(x) \,\mathrm{sh}\left(|\omega_{neg}|(t-t_0)\right)\;,
\end{equation}
where we fix $t_0=0$, $B_{neg}=10^{-2}$ in what follows.
Using the configuration~(\ref{eq:10}) and its time derivative as
Cauchy data at $t=0$, one solves numerically Eq.~(\ref{eq:1}) forward
and backward in time and obtains $\phi(t,x)$, see\footnote{Only
  the central parts of solutions are shown in this and other figures.}
Fig.~\ref{fig:sphaleron}b. The latter solution
interpolates between free wave packets 
above the vacuum $\phi_-$ and expanding bubble. We compute the values of
$(g^2E,g^2N)\approx (6.1,4.4)$ by Eqs.~(\ref{eq:7}) and mark the
respective point ``cb'' in Fig.~\ref{fig:sampling}a.
\FIGURE{\centerline{\includegraphics[width=0.5\textwidth]{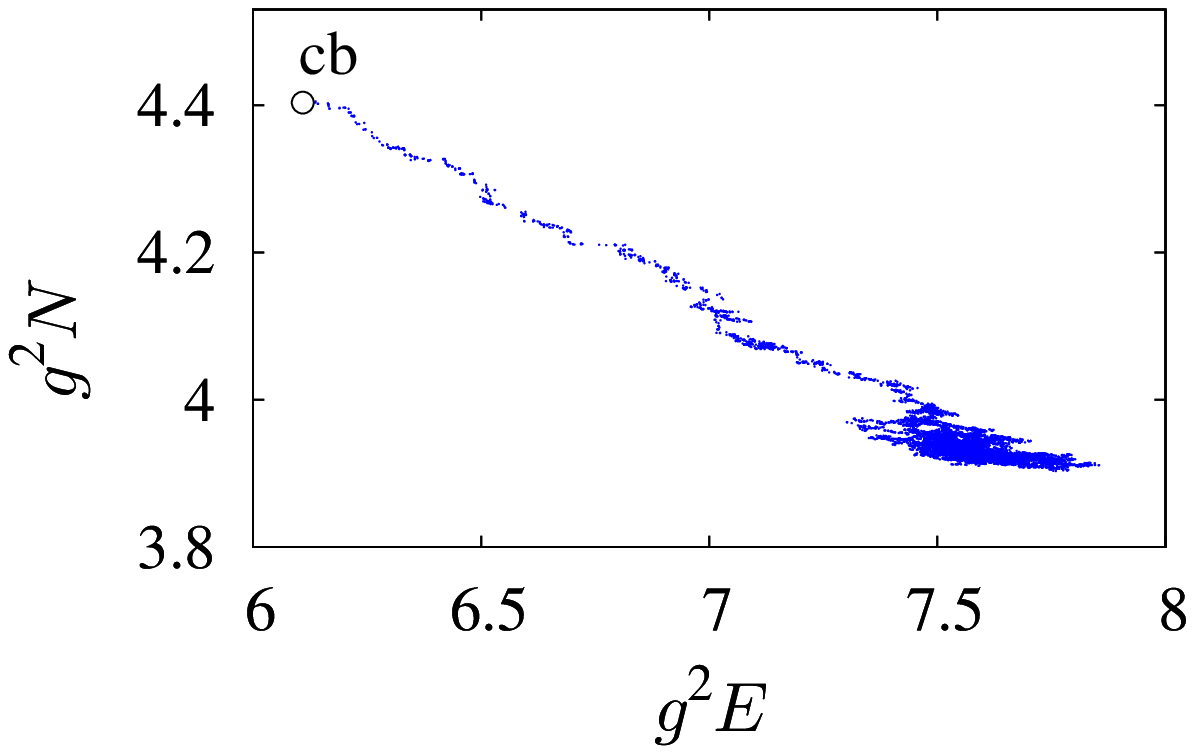}
\includegraphics[width=0.5\textwidth]{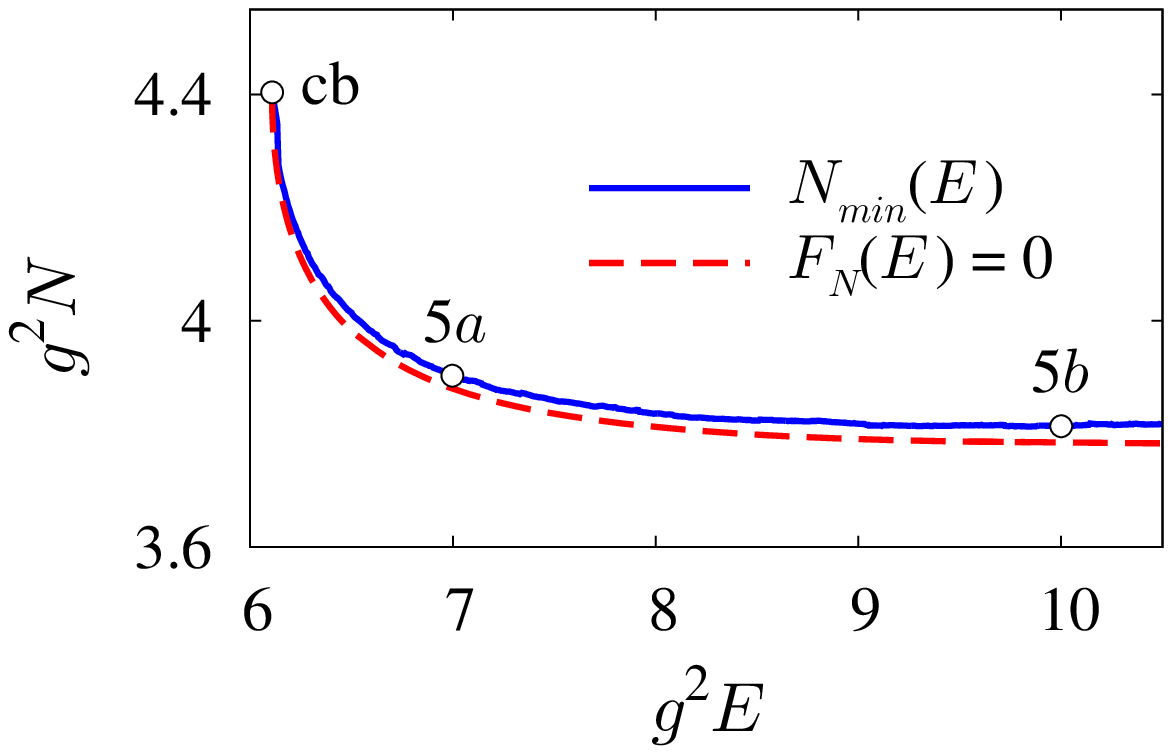}}
\centerline{\hspace{1.6cm}(a)\hspace{7.2cm}(b)}
\caption{\label{fig:sampling}(a) Stochastic sampling at $\tau=50$,
  $\vartheta=10^3$. (b) Lower boundary of the ``classically allowed''
  region. In both cases $\delta \rho=0.4$.}}

In Fig.~\ref{fig:sphaleron}b we check the linearity of evolution at $t\to
-\infty$ by comparing the linear and exact energies of $\phi(t,x)$,
Eqs.~(\ref{eq:7}) and (\ref{eq:8}). As expected, the linear energy
coincides with the exact one at large negative times and departs from
it when wave packets collide. In what follows we estimate the precision of
linearization as fractions of percent.

\subsection{Stochastic sampling technique}
\label{sec:stoch-sampl-techn}
In this Section we sample over classical solutions with bubbles
of true vacuum in the final state, see  Refs.~\cite{Rebbi1,Rebbi2}. It
is hard to pick up the initial Cauchy data for such solutions:
most of the initial wave packets scatter trivially and do not produce
expanding bubbles. Instead, we  consider the data
$\{\phi(0,x),\,\partial_t\phi(0,x)\}$ 
at $t=0$.  We decompose these data in the  basis of perturbations
around the critical bubble,
\begin{align}  
\label{eq:9}
&\phi(0,x) = \phi_{cb}(x) + A_{neg}\,\delta \phi_{neg}(x)  + \sum_k A_k\,
\delta\phi^{(cb)}_k(x)\;,\\  
\notag
&\partial_t\phi(0,x) = B_{neg}\,|\omega_{neg}|\,\delta \phi_{neg}(x) +
\sum_k B_k\, \omega_k^{(cb)}\,
\delta\phi^{(cb)}_k(x)\;,
\end{align}
where the negative mode is treated separately. Note that any
functions $\phi(0,x)$, $\partial_t\phi(0,x)$ can be written in the
form~(\ref{eq:9}). For each set of Cauchy data $\{A_{neg},\, 
A_k, $ $B_{neg},\, B_k\}$ we solve numerically Eq.~(\ref{eq:1}) and
obtain classical solution $\phi(t,x)$. Due to instability of the
critical bubble there is a good chance to obtain 
transition between the vacua $\phi_-$ and $\phi_+$. 

Next, we study the region in $(E,N)$ plane corresponding to
classical formation of expanding bubbles from colliding wave
packets. We are particularly interested in the lower boundary
$N = N_{min}(E)$ of this region. Let us organize the artificial ensemble
of solutions describing transitions between the vacua.  The
probability of finding each solution in our
ensemble is proportional to 
\begin{equation}
\label{eq:12}
p \propto \mathrm{e}^{-E\tau - N\vartheta}\;,
\end{equation}
where $E$ and $N$ are the energy and initial particle number of the
solution; $\tau$ and $\vartheta$ are fixed
numbers. At large positive $\vartheta$ solutions with the smallest $N$
dominate in the ensemble and we obtain 
the boundary $N_{min}(E)$ with good precision. Value of $\tau$ controls
the region of energies  to be covered.

We use Metropolis Monte Carlo algorithm to construct
the ensemble~(\ref{eq:12}). In our  approach solutions are
characterized by the coefficients in 
Eq.~(\ref{eq:9}); condition $A_{neg}=0$ is used to fix the time
translation invariance of Eq.~(\ref{eq:1}). The algorithm starts from
the solution (\ref{eq:10}) describing decay of the critical
bubble; it has 
$B_{neg} = 10^{-2}$, $A_k = B_k = 0$. Denote the energy and particle
number of this solution by $(E_0,N_0)$. We pick up a random coefficient
from the set $\{A_k,\, B_k,\, B_{neg}\}$ and change it by 
a  small step. The latter step is 
a Gauss--distributed random number with zero average and small
dispersion\footnote{We used $\sigma =  10^{-2},\, 10^{-3}$;
  the final result was insensitive to this number.}
$\sigma$. Substituting the modified set of coefficients into Eqs.~(\ref{eq:9}),
we find $\phi(0,x)$ and $\partial_t\phi(0,x)$. Then, solving numerically
the classical field  equation, we obtain the entire solution
$\phi(t,x)$. We compute the values of $(E, N)$ by 
Eqs.~(\ref{eq:7}). Solution is rejected if it does not interpolate
between the vacua $\phi_-$ and $\phi_+$; otherwise we accept it with
the probability 
\begin{equation}
\label{eq:13}
p_{accept} = \min \left( 1,\; \mathrm{e}^{-\tau\Delta E - \vartheta \Delta
N}\right)\;,
\end{equation}
where $(\Delta E,\Delta N)$ are differences between the new values of
$(E,N)$ and the values $(E_0,N_0)$ for the solution we started
with. If the new 
solution is accepted, we write it down and use its parameters $\{A_k,\,
B_k,\, B_{neg}\}$, $(E,N)\to (E_0,N_0)$ for the next cycle
of iterations. After many cycles we obtain the ensemble
(\ref{eq:12}) of accepted solutions.

A typical run of the Monte Carlo algorithm is shown in
Fig.~\ref{fig:sampling}a where the accepted solutions are marked by dots
in $(E,N)$ plane. The algorithm starts in the vicinity of the critical
bubble, then moves to smaller $N$ and finally arrives to the boundary
$N_{min}(E)$ where the majority of solutions is found.

\section{Numerical results}
\label{sec:results}
We perform Monte Carlo runs at different values of $\tau$ and $\vartheta$
until the entire curve $N=N_{\min}(E)$ is covered with solutions. In
total we
obtained $2\cdot 10^7$ solutions, where the value of $\tau$ was ranging
between $0$ and $10^4$; $\vartheta = 10^3,\, 10^4,\, 5\cdot10^4$.

The boundary $N=N_{min}(E)$ is constructed by breaking the energy
range into small intervals $\Delta E = 0.01$ and choosing solution
with minimal $N$ inside each interval. This is the result we are
looking for: $N_{min}(E)$ gives the minimum number of particles
needed for classically allowed production of bubbles. It is plotted in
Fig.~\ref{fig:sampling}b, solid line. As expected, $N_{min}(E)$ starts
from $(E,N)=(E_{cb},N_{cb})$ and decreases monotonously with
energy. At high energies $N_{min}(E)$ is approximately constant. Note
that the particle number is parametrically large in the ``classically
allowed'' region, ${N_{min} \sim 1/g^2}$. This means, in particular,
that the probability of producing the bubble from few--particle
initial states is exponentially suppressed. 

Given the boundary $N_{min}(E)$, we check results of Ref.~\cite{DL}
where classically forbidden transitions between $N$--particle states
and states containing the bubble were considered. The probability of
these processes is exponentially suppressed in the semiclassical
parameter,
\begin{equation}
\label{eq:14}
{\cal P}_N(E) \sim \mathrm{e}^{-F_N(E)/g^2}\;,
\end{equation}
where $F_N(E)$ is suppression exponent. One expects that this exponent 
vanishes in the 
``classical'' region  ${N>N_{min}(E)}$. We extract the boundary of
the set 
$F_N(E)=0$ from the results of Ref.~\cite{DL} and plot this boundary in
Fig.~\ref{fig:sampling}b (dashed line). It coincides with
$N_{min}(E)$ within 0.5\% accuracy; the agreement justifies both
calculations.

Let us look at solutions with almost--minimal initial particle number,
$N\approx N_{min}(E)$. Two such solutions are plotted\footnote{Small
  ripples covering the solutions represent fluctuations due to stochastic
  sampling.} in Fig.~\ref{fig:solutions}, their parameters
$(E,N)$ are shown by circles in Fig.~\ref{fig:sampling}b.
\FIGURE{
\unitlength=0.01\textwidth
\begin{picture}(100,40)
\put(4,5){\includegraphics[width=0.45\textwidth]{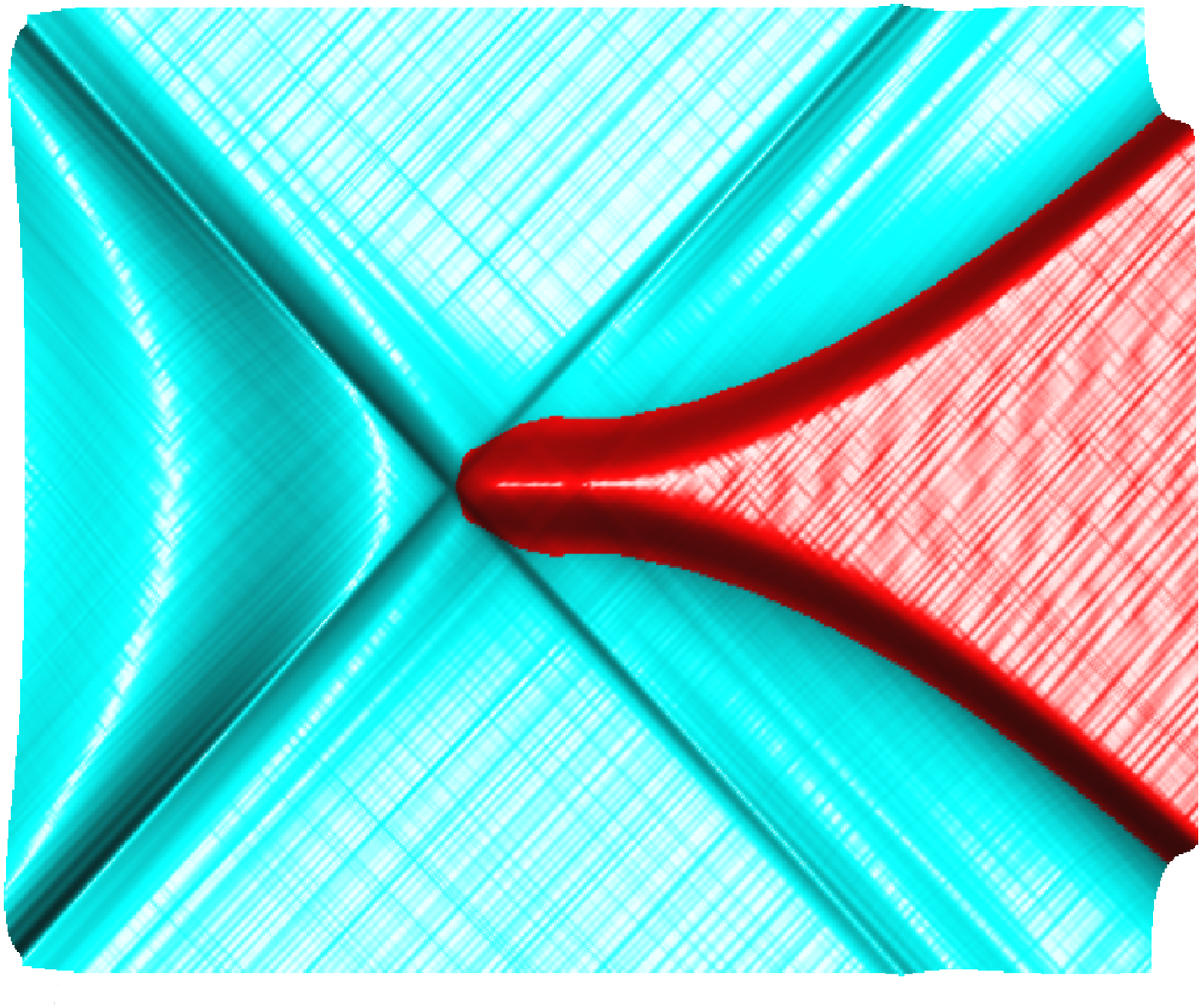}}
\put(55,5){\includegraphics[width=0.45\textwidth]{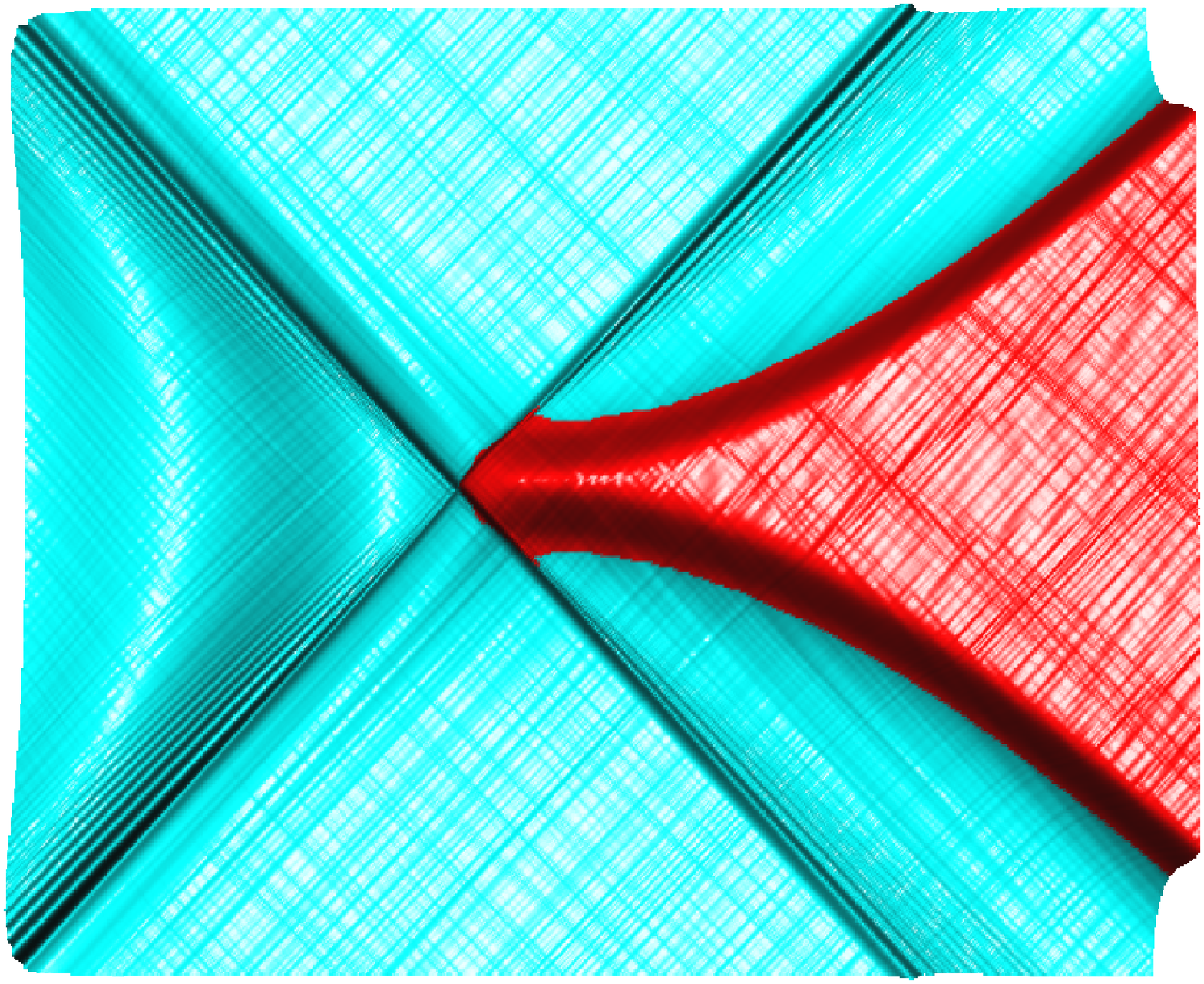}}
\put(76,0){\large $t$}
\put(25,0){\large $t$}
\put(51.5,23.5){\large\begin{sideways}$x$\end{sideways}}
\put(0,23.5){\large\begin{sideways}$x$\end{sideways}}
\put(55.5,36.9){\footnotesize $6$}
\put(55.5,32.3){\footnotesize $4$}
\put(55.5,27.8){\footnotesize $2$}
\put(55.5,23.2){\footnotesize $0$}
\put(53.5,18.6){\footnotesize $-2$}
\put(53.5,14.1){\footnotesize $-4$}
\put(53.5,9.5){\footnotesize $-6$}
\put(4.5,36.9){\footnotesize $6$}
\put(4.5,32.3){\footnotesize $4$}
\put(4.5,27.8){\footnotesize $2$}
\put(4.5,23.2){\footnotesize $0$}
\put(2.5,18.6){\footnotesize $-2$}
\put(2.5,14.1){\footnotesize $-4$}
\put(2.5,9.5){\footnotesize $-6$}
\put(96,4){\footnotesize $10$}
\put(92.3,4){\footnotesize $8$}
\put(88.6,4){\footnotesize $6$}
\put(84.4,4){\footnotesize $4$}
\put(80.2,4){\footnotesize $2$}
\put(76.0,4){\footnotesize $0$}
\put(70.4,4){\footnotesize $-2$}
\put(66.2,4){\footnotesize $-4$}
\put(62.0,4){\footnotesize $-6$}
\put(57.8,4){\footnotesize $-8$}
\put(45,4){\footnotesize $10$}
\put(41.3,4){\footnotesize $8$}
\put(37.6,4){\footnotesize $6$}
\put(33.4,4){\footnotesize $4$}
\put(29.2,4){\footnotesize $2$}
\put(25.0,4){\footnotesize $0$}
\put(19.4,4){\footnotesize $-2$}
\put(15.2,4){\footnotesize $-4$}
\put(11.0,4){\footnotesize $-6$}
\put(6.8,4){\footnotesize $-8$}
\end{picture}
\centerline{\hspace{0.6cm}(a) \hspace{6.9cm} (b)\hspace{3mm}}
\caption{\label{fig:solutions}Solutions $\phi(t,x)$ at $\delta
  \rho=0.4$. (a) $(g^2E,\, g^2N) \approx 
  (7,\, 3.9)$, (b) $(g^2E,\,g^2N) \approx (10,\, 3.8)$.}}
 At $t\to -\infty$ the solutions
describe free wave packets moving in the vacuum $\phi_-$. After collision 
the wave packets emit waves and form the bubble.  

The most surprising part of the evolutions in Fig.~\ref{fig:solutions} is
emission of waves during the bubble formation. One assumes that the
role of these waves is simply to carry away the energy excess which
is not required  for the creation of bubble. Indeed, solutions at
different $E$ look alike, 
cf. Figs.~\ref{fig:solutions}a and~\ref{fig:solutions}b; besides,
$N_{min}(E)$ is independent of energy at high values of the latter.
In Refs.~\cite{Shifman,Levkov,Levkov1} it was assumed that there exists
certain  limiting energy $E_{l}$ which is best for bubble creation. 
Then, classical solutions with 
minimal particle number at $E > E_{l}$  are sums of two parts:
non--trivial soft part describing bubble production at $E = E_{l}$ and
trivial hard part --- waves propagating 
adiabatically in the soft background. Hard waves carry away the energy 
excess $E - E_l$ without changing the initial particle number; 
this is achieved at small wave amplitudes and high frequencies, see
Eqs.~(\ref{eq:7}).  

Numerical results do not permit us to judge whether the limiting 
energy exists. We can, however, confirm the conjectured structure
of high--energy solutions. Consider the energies
$\epsilon_k=\omega_k |a_k|^2/g^2$ of the modes at $t\to -\infty$, where the
amplitudes $a_k$ and frequencies $\omega_k$ are defined in
Eq.~(\ref{eq:6}). In 
Fig.~\ref{fig:fk} we plot these energies for the two solutions
depicted in
Fig.~\ref{fig:solutions}. 
\FIGURE{\centerline{\includegraphics[width=0.6\textwidth]{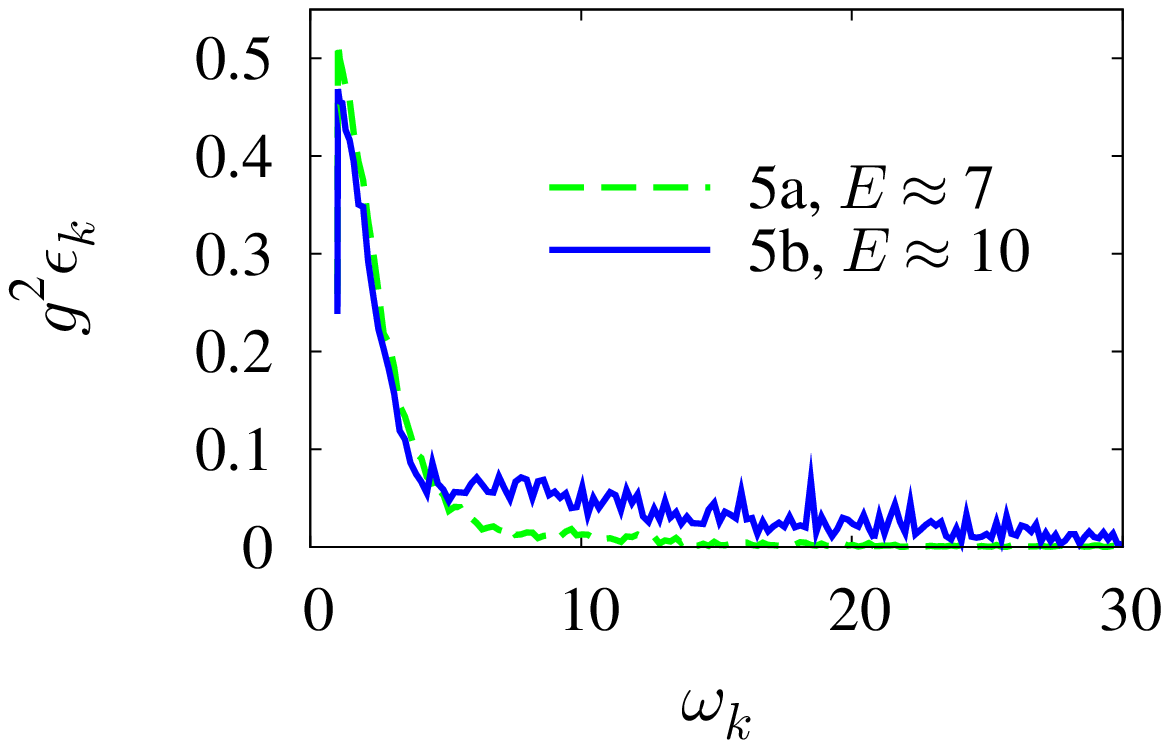}}  
  \caption{\label{fig:fk}Mode energies at $t\to -\infty$ for the
  solutions in Fig.~\ref{fig:solutions}.}}
Soft parts of the graphs are almost coincident. Long tail of excited
high--frequency modes is seen, however, in the graph representing the
high--energy 
solution~\ref{fig:solutions}b. The tail carries substantial energy
while the  
corresponding modes propagate adiabatically and do not participate in
nonlinear dynamics. This is precisely the behavior conjectured in
Refs.~\cite{Shifman,Levkov,Levkov1}.

It is reasonable to assume that solutions at arbitrarily high energies
have the same ``hard+soft'' structure. Then, $N_{min}(E)$ stays
constant as $E\to +\infty$ and classical formation of
bubbles is  not possible at any energies unless the 
initial particle number is larger than $N_{min}(E=+\infty)$.

Finally, we consider the limit ${\delta\rho \to 0}$. A typical solution at
small $\delta \rho$ is depicted in Fig.~\ref{fig:drho}a.
\FIGURE{
\unitlength=0.01\textwidth
\begin{picture}(100,40)
\put(50,1){\includegraphics[width=0.5\textwidth]{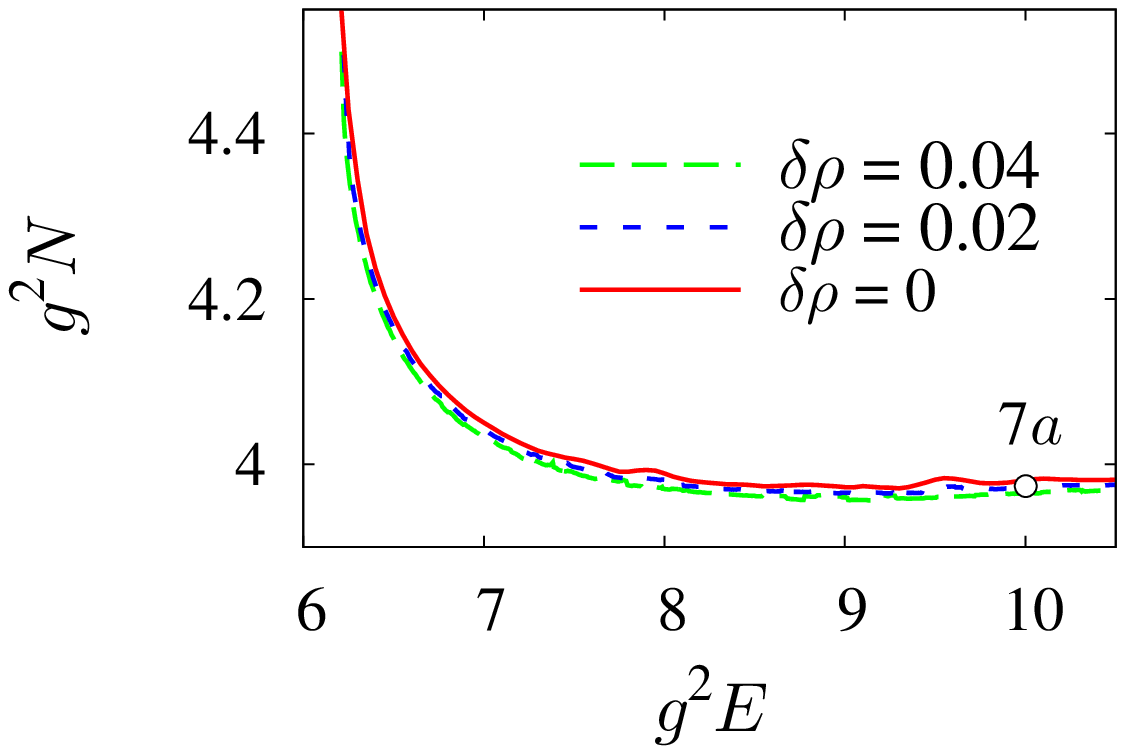}}
\put(5,5){\includegraphics[width=0.45\textwidth]{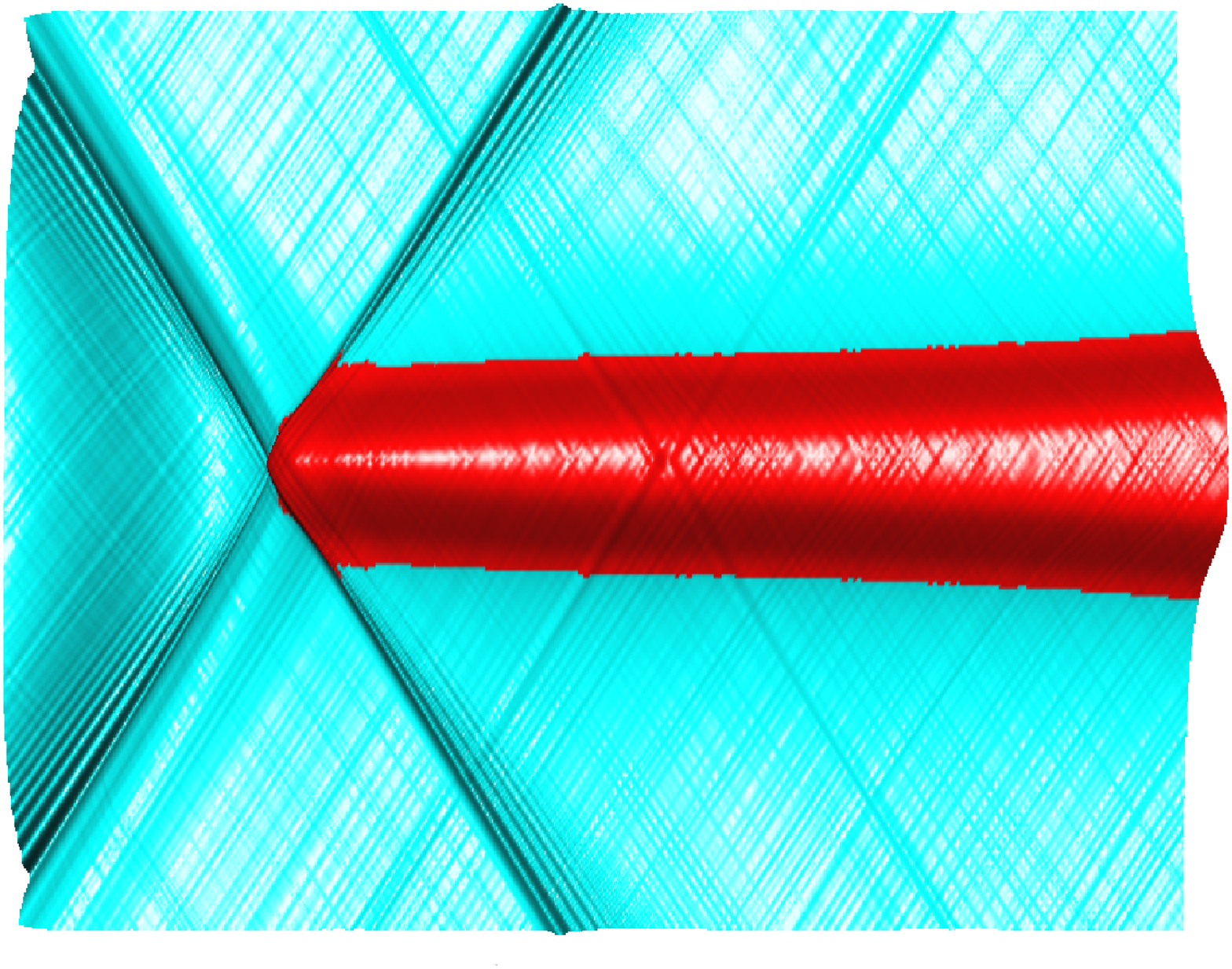}}
\put(28,0){\large $t$}
\put(0,21.8){\large\begin{sideways}$x$\end{sideways}}
\put(4.5,34.2){\footnotesize $4$}
\put(4.5,28.0){\footnotesize $2$}
\put(4.5,21.9){\footnotesize $0$}
\put(2.5,15.7){\footnotesize $-2$}
\put(2.5,9.5){\footnotesize $-4$}
\put(46,4){\footnotesize $10$}
\put(37.4,4){\footnotesize $5$}
\put(27.9,4){\footnotesize $0$}
\put(16.7,4){\footnotesize $-5$}
\put(6,4){\footnotesize $-10$}
\end{picture}
\centerline{\hspace{1cm}(a)\hspace{7.1cm}(b)}
\caption{\label{fig:drho}(a) Solution at $\delta\rho=0.02$,
  $(g^2 E,g^2 N)\approx (10,\, 4)$. (b) Lower
  boundary $N_{min}(E)$ at $\delta\rho\to 0$.}}
It describes creation of soliton and antisoliton which move away from
each other at
a constant speed. The boundaries $N_{min}(E)$ at
$\delta\rho=0.04,\; 0.02$ are plotted in Fig.~\ref{fig:drho}b (dashed
lines).  They are almost indistinguishable; thus, the 
limit ${\delta\rho \to 0}$ exists. Extrapolating $N_{min}(E)$ to $\delta
\rho = 0$ with linear functions, we obtain the region in $(E,N)$ plane
for classically allowed production of
soliton--antisoliton pairs (above the solid line in
Fig.~\ref{fig:drho}b). All initial wave packets leading to classical
creation of soliton pairs have the values of $(E,N)$ within
this region. The region in Fig.~\ref{fig:drho}b is qualitatively
similar to the regions at $\delta\rho>0$; in particular, $N_{min}(E)$
is constant at high energies.

\section{Discussion}
\label{sec:discussion}
In this paper we studied multiparticle states leading to classically
allowed production of soliton--antisoliton pairs in
$(1+1)$--dimensional scalar field model. We characterized
these 
states with two parameters  --- energy $E$ and particle number $N$; we
have found the corresponding ``classically allowed'' region in $(E,N)$
plane. There were two  
main ingredients in our technique. First, we added constant pressure
$\delta\rho$ pulling soliton and antisoliton apart. This modification
led to appearance of the critical 
\begin{wrapfigure}{r}{2cm}
\unitlength=1mm
\begin{picture}(25,65)
\put(0,5){\includegraphics[width=15mm]{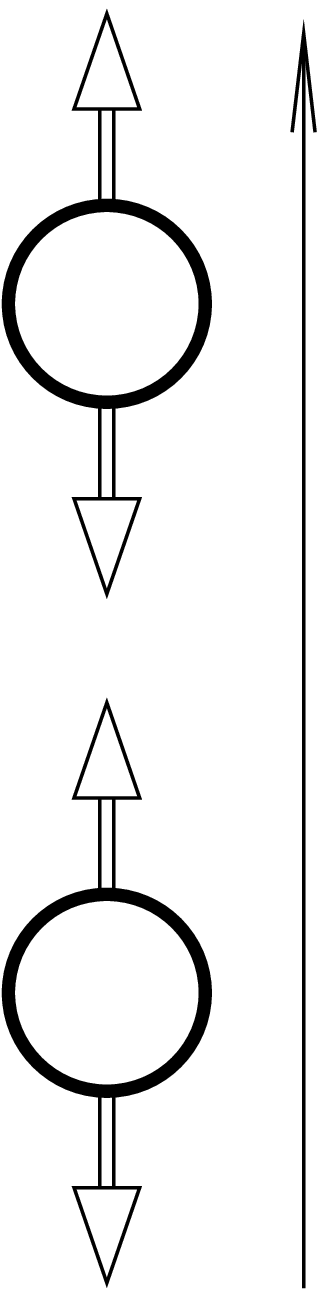}}
\put(2,16.5){\Large $\bar{M}$}
\put(2,49){\Large $M$}
\put(1,34){$\boldsymbol{F}_{att}$}
\put(-2,1.2){$-g_m\boldsymbol{H}$}
\put(0.3,65.5){$g_m\boldsymbol{H}$}
\put(16,61){$\boldsymbol{H}$}
\end{picture}
\end{wrapfigure}
bubble --- unstable static solution
lying on the boundary between the perturbative states and
soliton--antisoliton pair. Second, we applied  stochastic sampling
over Cauchy data in the background of the critical bubble. We thus
obtained large ensemble of classical solutions describing formation of
soliton--antisoliton pairs. Calculating the values of $(E,N)$ for each
solution we found the  required ``classically allowed'' region. 

Our method is naturally generalized to higher--dimensional models. For
example, consider formation of t'Hooft--Polyakov
monopole--antimonopole pairs in four--dimensional gauge
theories~\cite{Hooft,Polyakov}. Constant force dragging monopole and
antimonopole apart is provided~\cite{Shnir_Kiselev} by external
magnetic field $\mathbf{H}$, see the figure. 
At
$\mathbf{H}\ne 0$ there exists a direct analog of the critical
bubble~\cite{Manton}: unstable static solution where the attractive force
$\boldsymbol{F}_{att}$ between the monopole and antimonopole is
compensated by the external forces $\pm g_m\boldsymbol{H}$ ($g_m$ is a
magnetic charge of the monopole). One performs Monte Carlo simulation
in the background of this static solution and obtains many classical
evolutions  between free wave packets and monopole--antimonopole
pairs. 

A particularly interesting application of our technique might be found
in the study of kink--antikink production in $(1+1)$--dimensional
$\phi^4$ theory. In this model the boundary $N_{min}(E)$ is
lowered~\cite{kinks_particles,Shnir} due to chaos. We do not expect
any difficulties related to nontrivial dynamics of solitons. However,
classification of initial data for kink--antikink formation may
require modification of our method.

\paragraph*{Acknowledgments.}
We are indebted to V.Y.~Petrov and I.I.~Tkachev for helpful
discussions. This work was supported in part by grants
NS-5525.2010.2, MK-7748.2010.2 (D.L.), RFBR-11-02-01528-a (S.D.),
the Fellowship of the
``Dynasty'' Foundation  (awarded by the Scientific board of ICPFM)
(D.L.) and Russian state contracts 02.740.11.0244, P520 (S.D.), P2598
(S.D.). Numerical calculations have been performed on 
Computational cluster of the Theoretical division of INR RAS.

\end{document}